\documentclass{elsart}
\usepackage{graphicx,amssymb}

\begin{document}

\begin{frontmatter}

\title{Metastable States of the Classical Inertial
Infinite-Range-Interaction Heisenberg Ferromagnet: 
Role of Initial Conditions}

\vskip \baselineskip

\author[rio,natal]{Fernando D. Nobre} and 
\ead{nobre@dfte.ufrn.br}
\author[rio]{Constantino Tsallis} \ead{tsallis@cbpf.br}

\address[rio]{
Centro Brasileiro de Pesquisas F\'{\i}sicas \\
Rua Xavier Sigaud 150 \\
22290-180 \hspace{5mm} Rio de Janeiro - RJ \hspace{5mm} Brazil}

\address[natal]{
Departamento de F\'{\i}sica Te\'orica e Experimental \\
Universidade Federal do Rio Grande do Norte \\
Campus Universit\'{a}rio -- Caixa Postal 1641 \\
59072-970 \hspace{5mm} Natal - RN \hspace{5mm} Brazil}

\begin{abstract}
A system of $N$ classical Heisenberg-like rotators, characterized by
infinite-range ferromagnetic interactions, is studied numerically within
the microcanonical ensemble through a molecular-dynamics approach. 
Such a model, known
as the classical inertial infinite-range-interaction 
Heisenberg ferromagnet, exhibits a
second-order phase transition within the standard canonical-ensemble
solution. The present numerical analysis, which is restricted to an energy
density slightly below criticality, compares the effects of different
initial conditions for the orientations of the classical rotators. 
By monitoring the 
time evolution of the kinetic temperature, we observe that the system may
evolve into a metastable state (whose duration increases linearly with $N$),
in both cases of maximal and zero
initial magnetization, before attaining a second plateau at longer times. 
Since the kinetic temperatures associated with the second plateau, in the
above-mentioned cases, do not coincide, the system may present
a three-plateaux (or even more complicated) structure for finite $N$.  
To our knowledge, this has never
before been observed on similar Hamiltonian models, such as the XY
version of the present model. 
It is also shown that the system is sensitive to the way that one 
breaks the symmetry of the 
paramagnetic state: different nonzero values for the initial magnetization 
may lead to sensibly distinct evolutions for the kinetic temperature,
whereas different situations with zero initial
magnetization all lead to the same structure. 
\end{abstract}


\end{frontmatter}

\section{Introduction}
Physical systems characterized by long-range interactions and/or long-range
microscopic memories have attracted the attention of many workers recently
\cite{abeokamoto,grigolini,gellmanntsallis}. The main motivation remains on
the fact that such systems present, during longstanding states,
inconsistencies with the  
standard Boltzmann-Gibbs (BG) statistical-mechanics formalism. As an
example, the ergodic hypothesis -- known to be a pillar of the
BG framework -- may be violated: a breakdown of
ergodicity, leading to a fractal (or even more complex) occupation of 
phase space, has been observed in some cases \cite{baldovin04}. 
Also, some thermodynamic
quantities -- expected to increase linearly with the size of the
system within the BG framework -- like the internal or free energies,
may exhibit a nonextensive behavior in these systems. It is becoming
evident that a new statistical formalism -- more general than the BG
framework -- should be used to describe such systems properly. Up to now,
the most successful proposal appears to be nonextensive
statistical mechanics \cite{abeokamoto,grigolini,gellmanntsallis}, 
based on a generalization of the BG
entropy as proposed in 1988 \cite{tsallis88}.   

Among many interesting systems exhibiting nonextensive behavior, 
special attention has been dedicated to a classical Hamiltonian system,
namely, the inertial long-range-interaction XY model, defined as  
an assembly of $N$ classical planar rotators interacting 
through a long-range potential 
\cite{antoni,latora98,anteneodo,latora99b,%
latora00b,latora01b,montemurro,campa00,campa01,%
latora02,bene,yamaguchicondmat}. 
In the case of infinite-range interactions, i.e., in the limit where the  
mean-field approach becomes exact for the thermal equilibrium state, 
a well-known continuous phase transition 
occurs. If one considers a
total energy close to and below the critical energy, there exists a basin
of attraction for the initial conditions for which the system gets
captured in a metastable state, whose duration increases with $N$, before 
attaining the terminal thermal equilibrium. Therefore, the particular
order that one considers for applying the two relevant limits of this 
problem, namely, the thermodynamic
($N \rightarrow \infty$) and the long-time ($t \rightarrow \infty$)
limits, is extremely pertinent. If one considers the 
thermodynamic limit before the long-time 
limit, the system will remain in the metastable state and will
never reach the terminal equilibrium state.  
Moreover, in such a metastable state,  
the maximum Lyapunov exponent approaches zero (consequently so does 
the whole
Lyapunov-exponent spectrum) \cite{anteneodo,latora01b,bene}, 
indicating that the system is not strongly chaotic.
These effects strongly suggest a breakdown of ergodicity, revealing 
that the phase space will possibly not be equally and completely covered 
in the infinite-time limit. 

As an extension of the above-mentioned system, the inertial classical
Heisenberg ferromagnet has been investigated recently \cite{nobre03}. Such
a system, which consists of a modification of the well-known
Heisenberg model, where the spins are replaced by classical rotators, was
considered in the limit of infinite-range interactions.
It was studied numerically within the microcanonical
ensemble, through a molecular-dynamics approach; the initial conditions
used for the spin variables correspond to maximal magnetization, 
i.e., all rotators aligned along a given direction. 
Such a system has shown to be even more intriguing than its XY
counterpart: the metastable state, observed in the corresponding XY model
only near criticality, occurs, in the Heisenberg case, for a whole range of
energies, which starts right below criticality and extends up to very high
energies \cite{nobre03}.
 
In the present work we investigate the role played by different initial
conditions for the spin variables on the dynamical behavior 
of the inertial classical 
infinite-range-interaction Heisenberg ferromagnet. 
It is shown that the dynamical evolution of the kinetic temperature is
directly related to the initial magnetization (consistently with what has 
been recently observed for the XY model 
\cite{pluchinocondmat1,pluchinocondmat2}).
In the next section we define the model and
the numerical procedure. In section 3 we present and discuss our results.

\section{The Model and Numerical Procedure}
The inertial classical infinite-range-interaction Heisenberg ferromagnet is
defined through the Hamiltonian

%\vspace{5mm}

\setcounter{enumi}{2}
\setcounter{enumii}{1}
\renewcommand{\theequation}{\arabic{enumi}.\arabic{enumii}}
\begin{eqnarray}
H & = & K+V = {1 \over 2} \sum_{i=1}^{N} \sum_{\mu} L_{i \mu}^{2}
+ {1 \over 2N} \sum_{i,j=1}^{N} (1 - \vec{S}_{i}.\vec{S}_{j})
\nonumber \\
& = & {1 \over 2} \sum_{i=1}^{N} \sum_{\mu} L_{i \mu}^{2}
+ {1 \over 2N} \sum_{i,j=1}^{N} \left(1 - \sum_{\mu}
S_{i \mu} S_{j \mu} \right), 
\end{eqnarray}

\vspace{5mm}

%\vskip \baselineskip
\noindent
where the index $\mu$ ($\mu = x,y,z$) denotes Cartesian 
components and $L_{i \mu}$ represents the $\mu$-component of the 
angular momentum (or the rotational velocity, since we are assuming 
unit inertial moments) of rotator $i$.
The rotators are allowed to vary
their directions continuously on a 
sphere of unit radius, leading to the constraint

\vspace{-5mm}

$$
\sum_{\mu} S_{i \mu}^{2} = 1 \qquad (i=1,2, \cdots N).
\eqno(2.2)
$$

%\vskip \baselineskip
\noindent
Due to the close analogy of the above model to the standard classical
Heisenberg ferromagnet, we shall, sometimes, refer to the one-dimensional
inertial constituents (rotators) as spin variables (see \cite{anteneodo} 
for a discussion about the presence of the factor $1/N$ in front of 
the potential term). 

The BG canonical-ensemble solution of the present model may be worked out
easily \cite{nobre03}. The internal energy per particle is given by

\vspace{-5mm}

$$
u = {1 \over \beta} + {1 \over 2} (1 - \vec{m}^{2}),
\eqno(2.3)
$$

%\vskip \baselineskip
\noindent
where $\beta =1/T$ (herein we work in units of $k_{B}=1)$. In the equation
above, $\vec{m}$ represents the magnetization per particle, whose modulus
may be calculated by solving the self-consistent equation

\vspace{-5mm}

$$
m \equiv |\vec{m}| = {I_{3/2}(\beta m) \over I_{1/2}(\beta m)}
= {\rm cotanh}(\beta m) - {1 \over \beta m},
\eqno(2.4)
$$

%\vskip \baselineskip
\noindent
with $I_{k}(y)$ denoting modified Bessel functions of the first kind 
of order $k$. 
This model exhibits a well-known continuous phase transition 
at $T_{c}=1/3$, i.e., $u_{c}=5/6$.   

The molecular dynamics follows from a direct integration of the equations
of motion

%\vspace{5mm}

\setcounter{enumi}{2}
\setcounter{enumii}{5}
\setcounter{equation}{0}
\renewcommand{\theequation}{\arabic{enumi}.\arabic{enumii}\alph{equation}}
\begin{eqnarray}
\dot{\vec{L}}_{i} & = & \vec{S}_{i} \times \left( {1 \over N} 
\sum_{j=1}^{N} \vec{S}_{j} \right) \qquad (i=1,2, \cdots , N),  \\
\dot{\vec{S}}_{i} & = & \vec{L}_{i} \times \vec{S}_{i}
\qquad (i=1,2, \cdots , N), 
\end{eqnarray}

\vspace{5mm}

%\vskip \baselineskip
\noindent
which correspond to a set of $6N$ equations to be handled numerically. 
For solving such a set of equations we have used a fourth-order
Runge-Kutta-Merson integrator 
\cite{lapidus} with a time step of 0.05, leading, respectively,
to the relative energy
and spin-normalization conservations of $10^{-4}$ and $10^{-3}$, 
or better. The total initial kinetic energy was divided
into three equal parts, each of them to be assigned 
to a given set of Cartesian components of angular velocities
$\{ L_{i \mu} \} \ (i=1,2, \cdots , N)$. We have always 
started the system with 
the so-called water-bag initial conditions 
\cite{latora01b,latora02} 
for each set of components of angular velocities, i.e.,  
each set $\{ L_{i \mu} \}$ was extracted from a symmetric uniform
distribution 
and then, 
translated and rescaled to have zero total momentum. 
In what concerns 
the spin variables, we have started our simulations with a certain
configuration, associated with a given magnetization $m(0)$ at time $t=0$. 
The initial spin configurations employed are described below.

\vskip \baselineskip

\noindent
1) Maximal magnetization ($m(0)=m_{z}(0)=1$): all spins aligned 
along the $z$-axis, which corresponds to zero initial potential energy.

\noindent
2) Spin directions chosen at random in the upper ($y,z$) semicircle ($z>0$):
$m_{x}(0)=m_{y}(0)=0$; $m_{z}(0)=2/\pi$.
 
\noindent
3) Spin directions chosen at random in the upper semisphere ($z>0$): 
$m_{x}(0)=m_{y}(0)=0$; $m_{z}(0)=1/2$.

\noindent
4) Zero magnetization: $m_{x}(0)=m_{y}(0)=m_{z}(0)=0$. There are several
possibilities for such an initial configuration, which differ from one
another in the spin-component dispersions. We have used four
different choices, listed below.

\noindent
I) The polar and azimuthal angles, associated with each spin variable,
chosen at random, $\theta \in [0,\pi]$, $\phi \in [0,2\pi]$. This
corresponds to 
$\langle S_{x}^{2} \rangle = \langle S_{y}^{2} \rangle = 1/4$; 
$\langle S_{z}^{2} \rangle = 1/2$. 

\noindent
II) Spin directions chosen at random in ($x,y$) plane. In this case one has 
$\langle S_{x}^{2} \rangle = \langle S_{y}^{2} \rangle = 1/2$; 
$\langle S_{z}^{2} \rangle = 0$.

\noindent
III) Spins uniaxially random ($z$ axis): 
$\langle S_{x}^{2} \rangle = \langle S_{y}^{2} \rangle = 0$; 
$\langle S_{z}^{2} \rangle = 1$.

\noindent
IV) Spherically-symmetric spin distribution: 
$\langle S_{x}^{2} \rangle = \langle S_{y}^{2} \rangle = 
\langle S_{z}^{2} \rangle = 1/3$.

%%%%%%%%%%%%%%%%%%%%%%%%%%%%%%%%%%%%%%%%%%%%%%%%%%%%%%%%%%%%%%%%%%%%%%%%%%%
\begin{figure}
\begin{center}
\includegraphics[height=9cm,width=9cm]{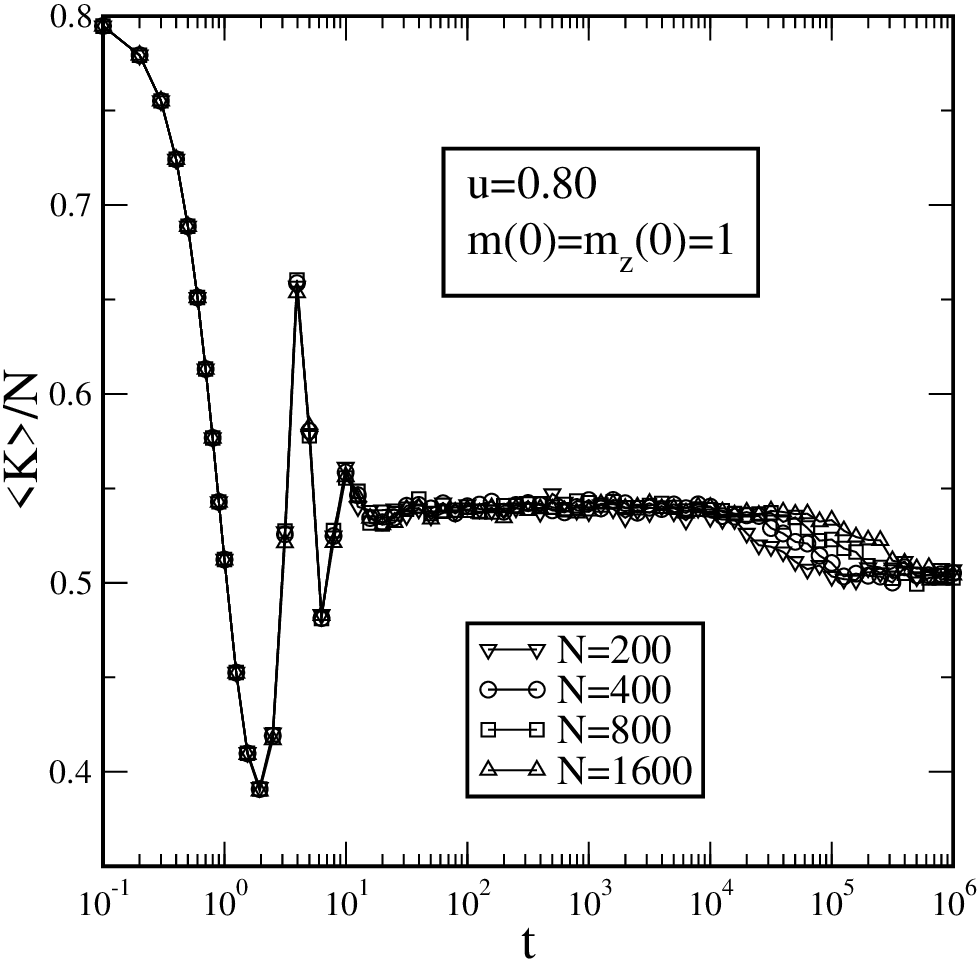}
\end{center}
\caption{\small
The microcanonical time evolution of $\langle K \rangle /N$ is represented
for several system sizes. 
The initial conditions are water-bag for velocities and maximal
magnetization for the spins. 
For the Hamiltonian defined in Eq. (2.1), energies are dimensionless
quantities. The time is also dimensionless and each unit of (physical) time
$t$ corresponds to 20 iterations of the equations of motion.} 
\label{fig1}
\end{figure}
%%%%%%%%%%%%%%%%%%%%%%%%%%%%%%%%%%%%%%%%%%%%%%%%%%%%%%%%%%%%%%%%%%%%%%%%%%%
\begin{figure}
\begin{center}
\includegraphics[height=9cm,width=9cm]{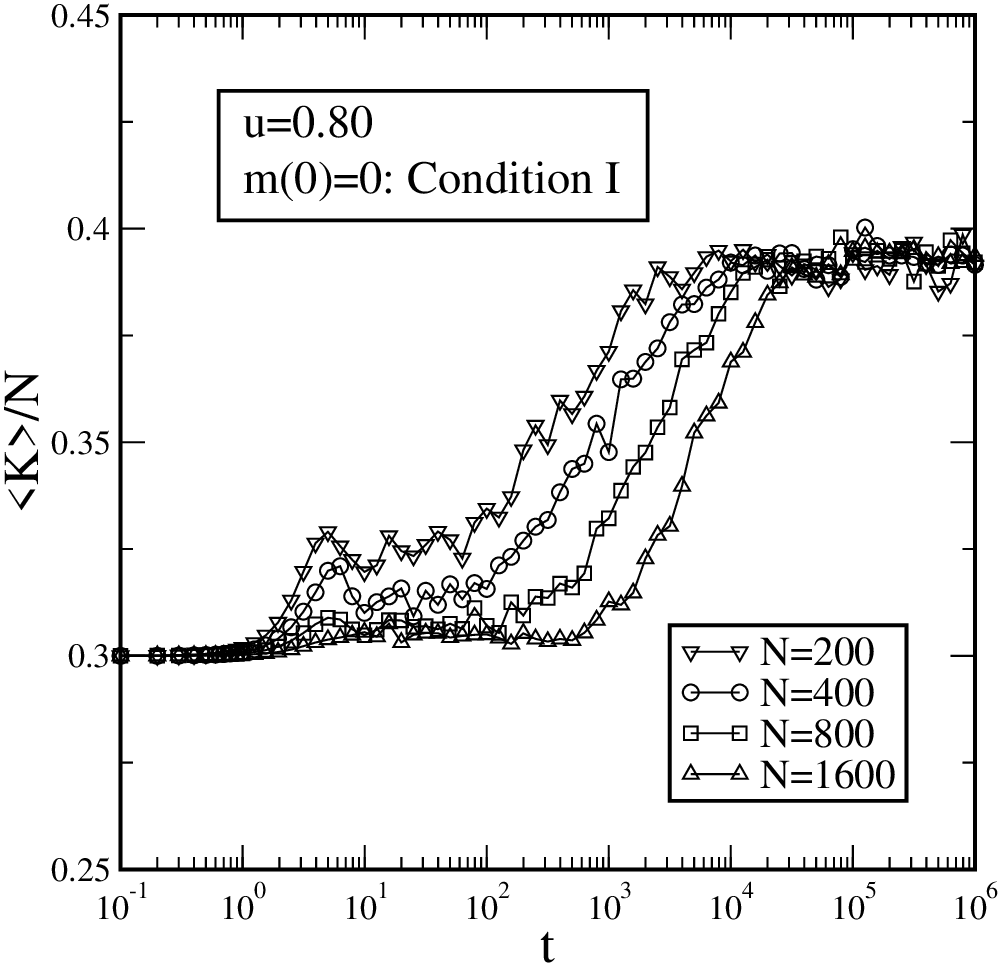}
\end{center}
\caption{\small
The microcanonical time evolution of $\langle K \rangle /N$ is represented
for several system sizes. 
The initial conditions are water-bag for velocities and zero magnetization
(condition I as described in the text) for the spins. 
For the Hamiltonian defined in Eq. (2.1), energies are dimensionless
quantities. The time is also dimensionless and each unit of (physical) time
$t$ corresponds to 20 iterations of the equations of motion.}
\label{fig2}
\end{figure}
%%%%%%%%%%%%%%%%%%%%%%%%%%%%%%%%%%%%%%%%%%%%%%%%%%%%%%%%%%%%%%%%%%%%%%%%%%%
\begin{figure}
\begin{center}
\includegraphics[height=9cm,width=9cm]{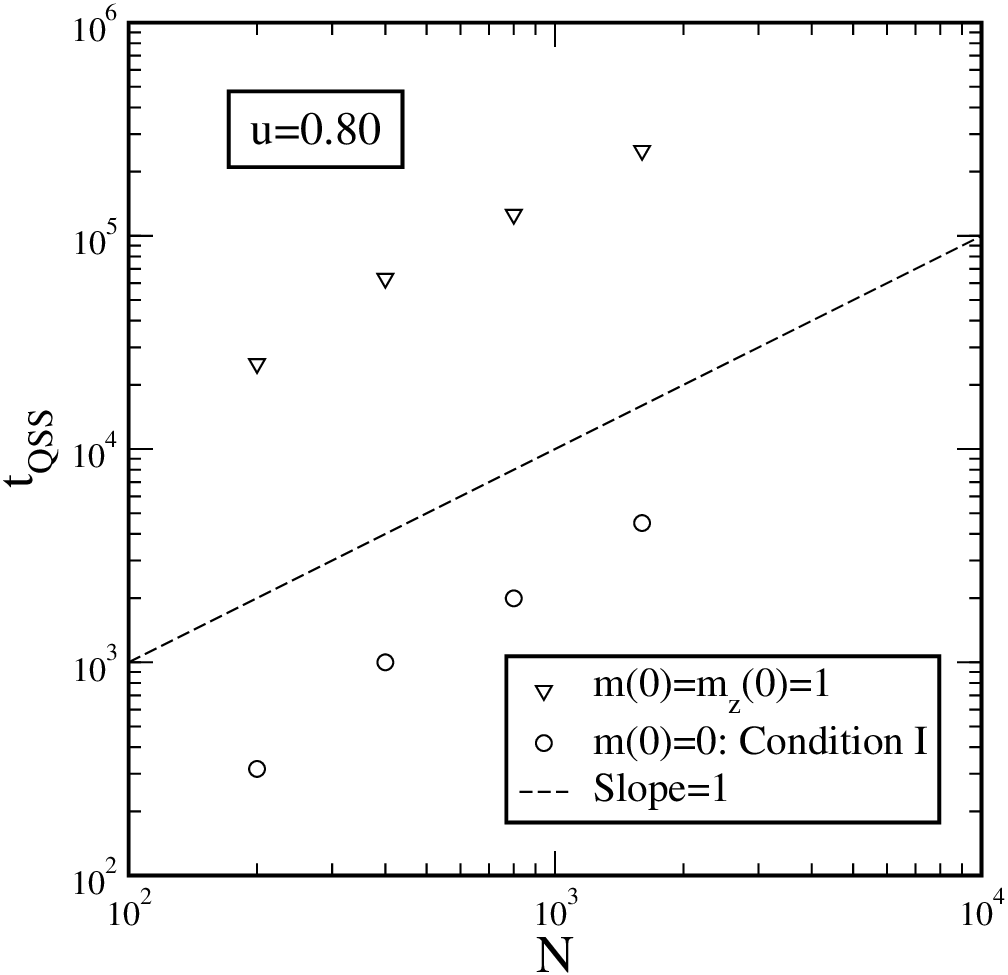}
\end{center}
\caption{\small
Log-log plots of the lifetime 
($t_{\rm QSS}$) of the QSS as a function of $N$, for the two cases 
considered in Figs. 1 and 2. In both cases, the slope is very close to 1
(represented by the dashed line), in such a way that $t_{\rm QSS} \sim N$.}
\label{fig3}
\end{figure}
%%%%%%%%%%%%%%%%%%%%%%%%%%%%%%%%%%%%%%%%%%%%%%%%%%%%%%%%%%%%%%%%%%%%%%%%%%%
\begin{figure}
\begin{center}
\includegraphics[height=9cm,width=9cm]{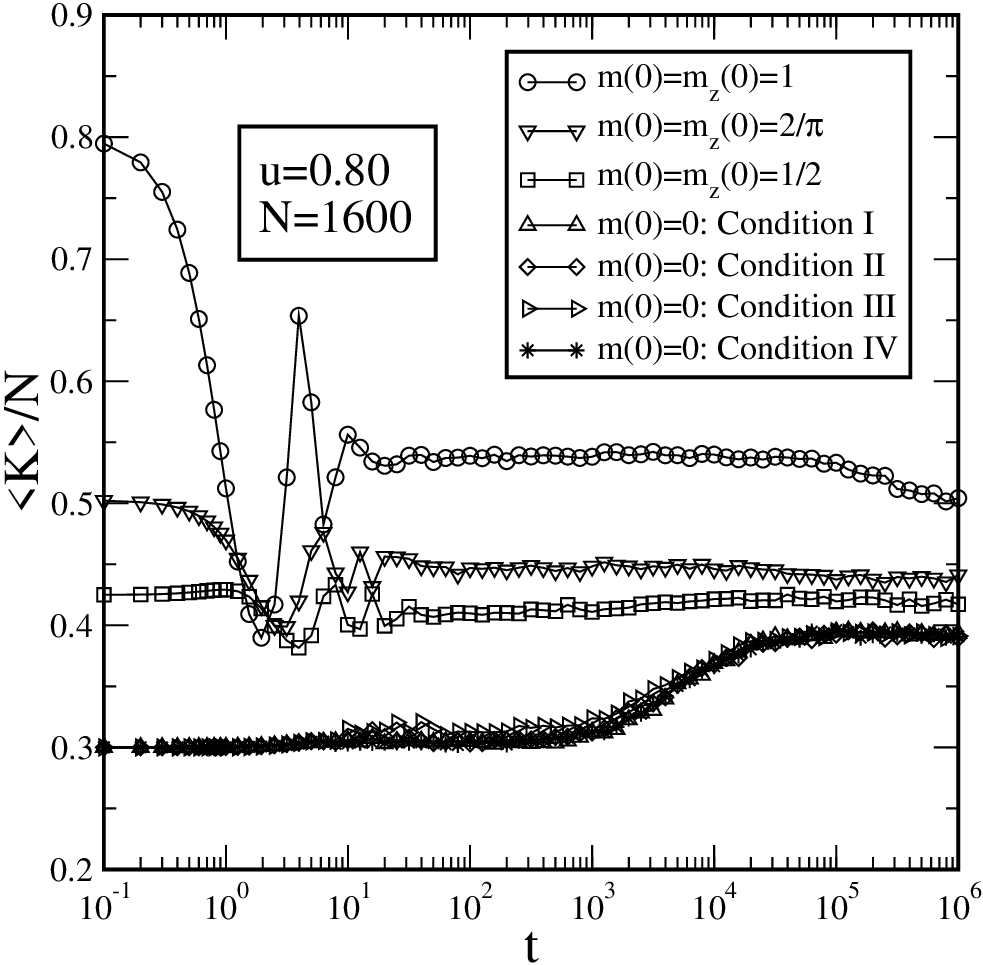}
\end{center}
\caption{\small
The microcanonical time evolution of $\langle K \rangle /N$ is represented
for different initial conditions of the spin variables (see text for a
description of such initial conditions). 
In all cases, the initial conditions for the velocities are water-bag.  
For the Hamiltonian defined in Eq. (2.1), energies are dimensionless
quantities. The time is also dimensionless and each unit of (physical) time
$t$ corresponds to 20 iterations of the equations of motion.} 
\label{fig4}
\end{figure}
%%%%%%%%%%%%%%%%%%%%%%%%%%%%%%%%%%%%%%%%%%%%%%%%%%%%%%%%%%%%%%%%%%%%%%%%%%%

\section{Results and Discussion}
In the results that follow, our measured quantities correspond 
to averages over $N_{s}$ distinct 
samples,
i.e., different initial sets of $\{ L_{i \mu} \}$ and $\{ S_{i \mu} \}$.
We have considered $N_{s}=20$ ($N=200$), $N_{s}=16$ ($N=400$),
$N_{s}=12$ ($N=800$), and $N_{s}=8$ ($N=1600$).
Our simulations were carried up to a maximum time $t_{\rm max}=10^{6}$ 
and each unit of 
(physical) time corresponds to 20 iterations of the equations of
motion. Herein, we restrict our analysis to an internal energy density
$u=0.8$, which is slightly below the critical energy
($u_{c}=5/6=0.8333...$). 
We have investigated, within our
microcanonical-ensemble molecular-dynamical approach, how 
$\langle K \rangle /N$ evolves in time (it should be mentioned that the
quantity $\langle K \rangle /N$, which represents an average over different
initial conditions of the kinetic energy per particle, when evaluated at
the $t \rightarrow \infty$ equilibrium, is expected to be proportional to
the temperature). 

In Fig. 1 we present the time evolution of $\langle K \rangle /N$, for
different values of $N$, in the
case of maximal magnetization. One observes that, after a short transient,
the system rapidly attains a metastable or quasistationary state (QSS), and
only after a long time 
does the system reach a second state, characterized by a lower value of 
$\langle K \rangle /N$. A similar effect is observed in the case of 
zero initial magnetization
(condition I as described above), as shown in Fig. 2. However, in the
second case, there is no short transient before the QSS, i.e., the system
is driven directly to the QSS at the initial time; in addition to that, 
the second state, obtained at longer times, presents a value of 
$\langle K \rangle /N$ that is higher than the one of the QSS. In both
cases, the lifetime of the QSS ($t_{\rm QSS}$) clearly increases 
with the size of the system. By defining $t_{\rm QSS}$ as the time at which
$\langle K \rangle /N$ presents its halfway between the values at the QSS
and the second state, one concludes that such a quantity asymptotically 
increases, linearly with $N$, i.e., $t_{\rm QSS} \sim N$ for the two cases, 
as shown in Fig. 3. As a consequence of this, 
for the cases considered in Figs. 1 and 2, if the
thermodynamic limit is performed before the long-time limit, the system
will remain in the QSS forever. 

In Fig. 4 we compare the time evolution of $\langle K \rangle /N$ for the 
various initial conditions of the spin variables described above (in all
cases, we have considered the size $N=1600$). One observes that this
property is not sensitive to the particular way one chooses for setting
a zero magnetization at the initial time: 
all four conditions corresponding to $m(0)=0$ lead to
the same time evolution of $\langle K \rangle /N$. However, each one of 
the three different choices of a nonzero $m(0)$ lead to 
a distinct time evolution of $\langle K \rangle /N$; this indicates that 
the time evolution of $\langle K \rangle /N$ is extremely dependent on the
particular value chosen for the initial magnetization. Up to the
maximal computational time ($t_{\rm max}=10^{6}$) considered in our
numerical analysis, we have observed a two-plateaux structure only
in the cases $m(0)=1$ and $m(0)=0$ (as shown in
Figs. 1--3). However, for the two cases $m(0)=2/\pi$ and $m(0)=1/2$, such
an effect is not so clear. An interesting point to stress is that 
the kinetic temperature is expected to coincide at an
infinite time for all cases considered in Fig. 4, i.e., the terminal
equilibrium should be independent of the particular initial conditions
employed. Therefore, if one considers the $t \rightarrow \infty$ before the
thermodynamic limit, there is a possibility of a three-plateaux (or even
more complicated) structure for some of the initial conditions considered
in Fig. 4.

>From the results above, one concludes that the inertial
infinite-range-interaction Heisenberg ferromagnet is even more intriguing
than its XY counterpart. For the latter one has, for energies slightly
below the critical energy, a QSS (whose lifetime diverges in the 
thermodynamic limit) followed by the corresponding terminal equilibrium
state; for
energies above the critical energy, such a QSS is not present (at least
in a clear way). A similar picture to that of the XY model close to
criticality occurs in the Heisenberg case for
energies considerably higher than the critical energy
\cite{nobre03}. Furthermore, close to criticality, the present 
Heisenberg model shows the possibility of a three-plateaux (or even more
complicated) structure, for a fixed system size. 
The picture that emerges, for explaining such an effect, 
is that the system may get trapped initially in a
small part of phase space; after some time, it will expand to a larger trap
(which may, presumably, contain the first one), and so
on, until it will finally explore the whole phase space.   
Obviously, a careful investigation of the properties of such a 
(possible) second QSS requires a high computational effort.  
However, if one considers the order of the
two relevant limits of this problem in such a way as to perform the 
thermodynamic
limit before the long-time limit, the system will remain in the first QSS
forever; all further states -- including other possible QSS's, as well as
the terminal equilibrium state -- 
will not be accessible to the system. In this case, the only relevant
state is the first QSS, which in the present Heisenberg model, 
may depend on the
particular initial conditions employed.

\section*{Acknowledgments}
We thank fruitful conversations with C. Anteneodo, E.~P. Borges,
E.~M.~F. Curado, A. Pluchino, and A. Rapisarda. 
The partial financial supports from
CNPq, Pronex/MCT and FAPERJ (Brazilian agencies) are acknowledged.
One of us (F.~D.~N.)
acknowledges CBPF (Centro Brasileiro de Pesquisas F\'{\i}sicas) for the 
warm
hospitality during a visiting period in which this work was accomplished.

%\newpage

\end{document}